\documentstyle[preprint,eqsecnum,aps]{revtex}
\begin{document}
\preprint{Penn Preprint UPR-0628T}
\preprint{WUGRAV 94-10}
\title{
High-frequency oscillations of Newton's constant
induced by Inflation
}
\author{Paul J. Steinhardt}
\address{
Department of Physics\\
University of Pennsylvania\\
Philadelphia, Pennsylvania 19104
}
\author{ Clifford M. Will}
\address{
McDonnell Center for the Space Sciences, Department of Physics\\
Washington University \\
St. Louis, Missouri 63130
}

\maketitle
\begin{abstract}
We examine the possibility that an epoch of inflationary expansion
induces high-frequency oscillations of Newton's constant, $G$.
The effect occurs because
inflation can shift the expectation value of a non-minimally
coupled, Brans-Dicke-like field away from the minimum of
its effective potential.
At some time after inflation ends,
the field begins to oscillate, resulting in
periodic variations in $G$.   We find conditions for which
the oscillation energy would be sufficient to close the universe,
consistent with  all known constraints
from cosmology and local tests of general relativity.
\end{abstract}


\section{Introduction and Summary}

Time-variation of Newton's constant, $G$, has been the subject of many
investigations.  Most studies, motivated by Dirac's large-number hypothesis,
focus on monotonic variations.  Brans-Dicke theory, for example, was devised as
a field theoretic model that incorporates a monotonic variation in $G$.
Its key element is a massless, scalar field, $\phi_{BD}$, which
couples non-minimally
to the scalar curvature, $\tilde R$,  through the interaction,
$\phi_{BD} {\tilde R}$.
The effective Newtonian constant is proportional to $\phi_{BD}^{-1}$.
The massless $\phi_{BD}$ increases uniformly in a matter dominated
epoch, resulting in a monotonic decrease in $G$.
The model is problematic from a quantum field theory point of view
because
there is no reason why the mass of $\phi_{BD}$ should remain zero after
quantum corrections.  The problem from an empirical point of view
is that there are tight limits from cosmology and local tests of
general relativity (for a review see \cite{tegp})

In this paper, we consider  the possibility of an {\it oscillating}
Newtonian constant in which $G$ varies
with a frequency
much greater than the expansion rate of the Universe,
$\nu \gg 10^{-17}$~Hz \cite{lowfreq}.
A
model with oscillations in $G$ is quite
simple to construct; {\it e.g.}, a massive Brans-Dicke model
in which a scalar field $\Phi$ (related to $\phi_{BD}$)
is displaced from the minimum of its effective
potential.  The displacement in $\Phi$ could result
naturally from an epoch of inflationary expansion.  During
inflation, the scalar curvature is non-zero, and so produces
a substantial contribution to the effective potential for $\Phi$
through the $\Phi R$ interaction.  The added contribution
shifts the minimum of the effective potential for $\Phi$.  After
reheating, the scalar curvature is zero, and $\Phi$ is free to
oscillate about the true minimum of its effective potential.  From
the point of view of quantum field theory, the massive
Brans-Dicke field does not suffer the naturalness problem of the
massless case.  One might  anticipate that the mass of $\Phi$
would be similar to that of typical elementary particles, $m \ge 1$~GeV
or so, a range corresponding to very high-frequency
($\nu^{-1} \sim m^{-1} << 1$~sec) oscillations.

Accetta and one of us \cite{Acc} have previously discussed
some  effects that
high-frequency oscillations of $G$ could have on cosmological measurements.
The purpose of the present study is to examine a very plausible scenario
in which inflation sets off the oscillations in $G$.
For simplicity, we consider cosmological models for which the
energy density contains
three components: ordinary, baryonic matter, $\rho_B$; radiation
$\rho_R$; and
oscillatory energy, $\rho_{\Phi}$.  It is straightforward to
extend our results to models which include a fourth,
conventional dark matter component.  We assume that some inflaton field
completely independent of $\Phi$ is responsible for providing the
vacuum energy density needed to drive inflation.  We further
assume that $\Phi$ is so weakly coupled that $\Phi$-particles
are not produced by the decay of the inflaton during reheating.
During inflation, the $\Phi$ field is shifted from the minimum
of its effective potential; after inflation,
the only effect on oscillations is the Hubble
red shift of the $\Phi$-field kinetic energy.
We find that the scenario passes
all known tests for  a wide range  of model
parameters.  In particular, it
is possible to find viable models in which the oscillatory energy in $\Phi$
provides all of the missing energy density needed to reach  the critical
density without introducing conventional dark matter.

The organization of the paper is as follows:  in Section II,
notation and the basic equations are introduced to describe
massive Brans-Dicke
cosmology with oscillating $G$.  In Sections III and IV,
 the oscillations in $G$ induced by an epoch of
inflationary expansion are computed  for  scalar fields
coupled to the scalar curvature through  linear and quadratic
interactions, respectively.  The red shifting of the oscillation
amplitude after inflation is analyzed to determine the
oscillation energy today.  It is shown that the oscillation
energy is compatible with all cosmological constraints for a
wide range of parameters and that it is possible for the
oscillation energy to reach closure density.  In Section V,
the constraints from local tests of general relativity are
examined, including measurements of
light deflection and time delay, data from laser ranging to Earth-orbiting
satellites and the Moon,  and tests of Newtonian gravity using tower and
laboratory experiments.  Section VI presents concluding remarks.

\section{Basic equations}

\subsection{Action and field equations}

The gravitational theory underlying the oscillating-$G$ cosmologies is
a general scalar-tensor metric gravitational theory, in which the
scalar field has both a mass and a non-linear $\lambda \phi^4$
self-interaction.  The action has the general form

\begin{equation}
I=\int \sqrt{-\tilde g} d^4x [ M^2 f(\phi) \tilde R - {1 \over 2} \nabla_\mu
\phi \nabla^\mu \phi - \tilde V (\phi) ] + I_m (\psi_m, \tilde
g_{\mu\nu})  , \label{action}
\end{equation}
where $M^2=(16\pi G)^{-1}$, $\tilde g_{\mu\nu}$ is the physical
metric, $I_m$ is the matter action, with matter fields $\psi_m$ coupling only
to $\tilde g_{\mu\nu}$ (hence a metric theory), and $\tilde R$ is the Ricci
scalar constructed from $\tilde g_{\mu\nu}$.  We use the standard
notation and conventions of \cite{tegp,MTW}.
The potential $\tilde V$ is given by
\begin{equation}
\tilde V(\phi)={1 \over 2} m^2 \phi^2 + {1 \over 4} \lambda \phi^4 \,.
\label{potential}
\end{equation}
With $\tilde V=0$, this would correspond to a generalized Brans-Dicke
(BD)
theory \cite{tegp} with $\phi_{BD}\equiv f(\phi)$ and the BD coupling
function
\begin{equation}
\omega=f/2(Mf^\prime)^2 ,
\label{omega}
\end{equation}
where ``prime'' denotes $d/d\phi$.  The field equations are
\begin{mathletters}
\label{allfieldeq}
\begin{equation}
\tilde G_{\mu\nu}= {1 \over {2M^2 f}} \tilde T_{\mu\nu} +
{1 \over f} (f_{;\mu\nu} -
\tilde g_{\mu\nu} \stackrel{\sim}{\Box} f )
+ {1 \over {2M^2f}} (\phi_{,\mu}\phi_{,\nu}- {1
\over 2} \tilde g_{\mu\nu}\phi_{,\lambda}\phi^{,\lambda} - \tilde
g_{\mu\nu} \tilde V(\phi)) ,
\label{fieldeq1}
\end{equation}
\begin{equation}
\stackrel{\sim}{\Box} \phi = {{fV^\prime -2f^\prime V - {1 \over 2}f^\prime
(1+6M^2f^{\prime\prime})\phi_{,\lambda}\phi^{,\lambda} + {1 \over
2}f^\prime \tilde T} \over {f+3M^2(f^\prime)^2}} ,
\label{fieldeq2}
\end{equation}
\end{mathletters}
where $\tilde T_{\mu\nu}$ is the physical stress-energy tensor derived
from the matter action $I_m (\psi_m,\tilde g_{\mu\nu})$, and
$\stackrel{\sim}{\Box}$
 is the d'Alembertian using the physical metric, and so on.

For cosmology, it turns out to be useful to recast the theory into a
representation with an Einstein gravitational action by making
a conformal transformation to the non-physical metric
\begin{equation}
g_{\mu\nu}\equiv f(\phi) \tilde g_{\mu\nu}.
\end{equation}
This representation is sometimes called the Einstein conformal frame
(see \cite{esposito} for discussion).
The action then takes the form
\begin{equation}
I=\int \sqrt{-g} d^4x [M^2R- {1 \over 2} \nabla_\mu \Phi \nabla^\mu
\Phi - V(\Phi)] + I_m (\psi_m, f^{-1}g_{\mu\nu}) ,
\label{action2}
\end{equation}
where $V(\Phi)=f^{-2} \tilde V (\phi)$ and $\Phi$ is defined by the
differential equation
\begin{equation}
d\Phi/d\phi=f^{-1}(f+3M^2{f^\prime}^2)^{1/2}.
\label{dphidphi}
\end{equation}
Damour and colleagues \cite{esposito,DEF,DN} have discussed general classes of
scalar-tensor theories in this framework; their scalar field $\varphi$
is given by $\varphi=\Phi/2M$, $A(\varphi)=f^{-1/2}$, and $V=0$.
We will consider models in which the non-minimal
coupling is linear, $f(\phi)=1+\alpha\phi/M$, and quadratic,
$f(\phi)=1+\xi\phi^2/M^2$.

The field equations in this representation are
\begin{mathletters}
\begin{equation}
G_{\mu\nu}= {1 \over {2M^2f}} \tilde T_{\mu\nu} + {1 \over {2M^2}}
(\Phi_{,\mu}\Phi_{,\nu}-{1 \over 2} g_{\mu\nu} \Phi_{,\lambda}\Phi^{,\lambda}
- g_{\mu\nu} V(\Phi) ) ,
\end{equation}
\begin{equation}
\Box \Phi = {d \over d\Phi} (V(\Phi) - {1 \over 4} f^{-2} \tilde T )
\,.
\end{equation}
\end{mathletters}
where now $\Box$ and $G_{\mu\nu}$ are defined using the non-physical
metric; however, $\tilde T_{\mu\nu}$ still refers to the physical
stress-energy tensor defined using locally measured quantities and the
physical metric.

\subsection{Cosmology}

Assuming a standard homogeneous and isotropic universe, we choose
coordinates $(t, \bf x)$ so that the ``line element'' constructed from
$g_{\mu\nu}$ has the Robertson-Walker form
\begin{equation}
ds^2=-dt^2+a(t)^2 d\sigma ^2 ,
\end{equation}
where $a(t)$ is the scale factor, and $d\sigma^2$ is the spatial line
element representing the hypersurfaces of homogeneity.  Since we will
assume a standard inflationary scenario for the early universe, we
will adopt a spatially flat metric, {\it i.e.} we will choose $k=0$.
The relation between $t$ and $a$ and the corresponding physical
variables $\tilde t$ and $\tilde a$  in the Robertson-Walker version
of the physical metric $\tilde g_{\mu\nu}$ is
$d \tilde t=f^{-1/2} dt$, $\tilde a =f^{-1/2} a$.  The physical
density and pressure $\rho$ and $p$ have their usual meanings as local
properties of the matter measured in local Lorentz frames of the
physical metric  $\tilde g_{\mu\nu}$.  Defining $H \equiv \dot a/a$,
where an overdot denotes derivative with respect to $t$,
we obtain the cosmological field equations and equations of motion
\begin{mathletters}
\begin{equation}
H^2=(6M^2)^{-1} \left ( \rho f^{-2} + {1 \over 2} {\dot\Phi}^2+V(\Phi)
\right ) \label{cosmo-a},
\end{equation}
\begin{equation}
\ddot \Phi+3H\dot\Phi=-dV(\Phi)/d\Phi-{1 \over 4}
(\rho-3p) d(f^{-2})/d\Phi \label{cosmo-b},
\end{equation}
\begin{equation}
\dot\rho=-3(\rho+p)(H-{1 \over 2}\dot f/f) \label{cosmo-c}.
\end{equation}
\end{mathletters}
Note that the use of the non-physical metric $g$ is the origin of the
$\rho f^{-2}$ and the $-{1 \over 2}\dot f/f$ terms in Eqs.
(\ref{cosmo-a}) and (\ref{cosmo-c}), respectively.
The physically measured Hubble parameter
$\tilde H = {\tilde a}^{-1} d \tilde a /d \tilde t$ is related to the
Hubble parameter used in the Einstein conformal representation by $\tilde H =
f^{1/2} (H-\dot f /f)$.

If $\rho-3p$ is $\Phi$-independent, as we will assume in this
paper, we can use Eq. (\ref{cosmo-b}) to define an effective potential
\begin{eqnarray}
V_{\rm eff} &\equiv & V(\Phi) + {1 \over 4}(\rho-3p)f^{-2}
\nonumber\\
&=&f^{-2}( \tilde V (\phi) + {1 \over 4}(\rho-3p) ) \,,
\label{Veff}
\end{eqnarray}
such that $\ddot \Phi +3H \dot \Phi = -dV_{\rm eff}/d\Phi$.

We assume that inflation is triggered by some inflaton field in the
matter sector (not $\Phi$!), whose vacuum energy density $\rho_V$
dominates the total energy density and leads to an epoch during which
$\rho_V=-p={\rm constant}$.  We relate the vacuum energy density to a
temperature by the standard relation (in units where $\hbar=c=k_B=1$)
\begin{equation}
\rho_V \equiv T_V^4 =(10^{14} {\rm GeV})^4 t_V^4
\label{rhov}
\end{equation}
where $t_V \equiv T_V/10^{14} {\rm GeV}$.  Note that, with these definitions,
$\rho_V/6M^4=2 \times 10^{-18} t_V^4$.

In characterizing cosmological evolutions, it will be useful to define
a number of key variables.

\noindent
(a) {\it Oscillation amplitude:}
\begin{equation}
\eta \equiv f(\phi)-1 = \cases{
\alpha\phi/M ,& [linear]\cr
\xi\phi^2/M^2 ,& [quadratic]\cr}
\label{eta}
\end{equation}

\noindent
(b) {\it Coupling parameter:}
\begin{equation}
\beta \equiv 3M^2{f^\prime}^2=\cases{
3\alpha^2,& [linear]\cr
12 \xi^2 \phi^2/M^2 ,& [quadratic]\cr}
\label{beta}
\end{equation}

\noindent
(c) {\it Anharmonicity parameter:}
\begin{equation}
\zeta \equiv \lambda\phi^2/2m^2 ,
\label{zeta}
\end{equation}

\noindent
(d) {\it Ratio of scalar-to-radiation energy density}
\begin{equation}
\chi \equiv \rho_\Phi /f^{-2}\rho_R \,,
\label{chi}
\end{equation}
where $\rho_R$ is the energy density of radiation,  and
\begin{equation}
\rho_\Phi = {1 \over 2} \dot \Phi^2 + V(\Phi) \,.
\label{scalarenergy}
\end{equation}
The parameter $\zeta$ determines the relative importance of the
$\lambda\phi^4$ anharmonic potential compared to the mass term
$m^2\phi^2$.
Note that $\beta$ is related to the Brans-Dicke parameter
$\omega$ by $\omega= {3 \over 2} (1+\eta)/\beta$; as
$\beta\to 0$ for finite $\eta$, $\omega \to \infty$, and the
theory becomes effectively general relativistic.

For future use we also define the dimensionless coefficients

\noindent
(e) Ratio of Planck mass to scalar mass:
\begin{equation}
y\equiv M/m = 1.7 \times 10^{18}(1 {\rm GeV}/m) \,   ,
\label{y}
\end{equation}

\noindent
(f) $Q$
\begin{equation}
Q \equiv 12 \lambda^{-1} (\rho_V/6M^4) = 2.4 \times 10^{-17} \lambda^{-1}
t_V^4 \,.
\label{Q}
\end{equation}
We assume that the material content of the universe is strictly that
of ordinary matter (no exotic dark matter), and we
will treat the equation of state for matter as having three phases,
inflationary, radiation dominated and matter dominated, with the ratio
$p/\rho \equiv \nu = \rm{constant}$ during each phase, with the respective
values -1, 1/3, and 0.

In all our models we will assume that the present amplitude of
$G$-oscillations is very small, {\it i.e.} $\eta_0 << 1$, where the
sub- or superscript $0$ denotes present values.  We define density
parameters at the present epoch in the Einstein conformal frame: for
matter (baryons plus radiation): ${\Omega_m}^0 = \rho^0/6M^2
H_0^2$;
for radiation alone: ${\Omega_R}^0 = \rho_R^0/6M^2
H_0^2$;
for the scalar field: ${\Omega_\Phi}^0={\rho_\Phi}^0 /6M^2 H_0^2$.
Because of the present smallness of $\eta$, and because
of its rapid oscillations compared to the expansion timescale, the
physically measured $\tilde H_0 \approx H_0$, so $\Omega_m^0$,
$\Omega_R^0$,
$\Omega_\Phi^0$ and $\chi_0$ are interchangable with the physically
measured quantities.  We note that
\begin{equation}
\Omega_R^0 = 4.3 \times 10^{-5} h^{-2} (T_0/2.73)^4 \,,
\label{omegarad}
\end{equation}
where $h$ is the present Hubble parameter in units of 100 km s$^{-1}$
Mpc$^{-1}$ ($0.4 < h<1$), and $T_0$ is the temperature of the cosmic
microwave background in degrees Kelvin.

Since $\Phi$ interacts only gravitationally , it is thermally
decoupled from ordinary matter; we ignore the possibility of particle
creation induced by the non-minimal coupling of $\phi$ to curvature.
Consequently, we
can relate the energy densities of baryons
$\rho_B$ and radiation $\rho_R$ to the scale factor $a$ and
temperature $T$ using the matter equation of motion in the physical
representation,
$d \rho /d \tilde t= -3(\rho+p) \tilde H$, combined
with standard thermal equilibrium
arguments.  Thus, when baryons and radiation are decoupled, we have
$\rho_B \propto {\tilde a}^{-3} $, $\rho_R \propto {\tilde a}^{-4} $, $T_R
\propto {\tilde a}^{-1}$.  Then, denoting the
end of inflation, marked by the conversion of
false vacuum energy density to radiation, by I, it is straightforward
to show \cite{kolbturner} that the non-physical scale factors at these
epochs are
related by
\begin{equation}
a_0/a_I = (T_I/T_0 )(f_0/f_I)^{1/2}
 \approx 4\times 10^{26} (f_0/f_I)^{1/2}(T_V/10^{14} {\rm GeV}) \,.
\label{scalefac}
\end{equation}

\subsection{Evolution of oscillating scalar field}

In the limit of small $\Phi$, the basic oscillation frequency of
$\Phi$ is $\sim m$.  Roughly speaking, in an expanding cosmology,
$\Phi$ will oscillate when the period of oscillation $\sim m^{-1}$  is
small compared to the expansion timescale $\sim H^{-1}$, {\it i.e.}
when $m>>H$.  Because the equation for $\Phi$ is non-linear, the
effective frequency of oscillation could differ substantially from
$m$.  As a consequence, we will find it necessary to give a refined
criterion for the onset of oscillations, separately applicable for the
linear and quadratic models which we are considering.  Once
oscillations commence, however, the subsequent evolution of the
oscillation amplitude can be found using a generalization of the
method used by Turner $\cite{turner}$.

We first convert the field equation for $\Phi$ into an approximate
``energy conservation'' equation, by multiplying Eq. (\ref{cosmo-b}) by
$f^n \dot \Phi$ and extracting total time derivatives.  The result is
\begin{equation}
{d \over {dt}} [ f^n({1 \over 2} {\dot
\Phi}^2+V(\Phi))]=-3Hf^n{\dot\Phi}^2+6M^2H^2 d f^n /dt \label{edot},
\end{equation}
where $n={1 \over 2} (1-3\nu)=2 \,,0\,, {1 \over 2}$, for inflation,
radiation domination and matter domination, respectively, and where we
assume that $n$ is constant during each phase dominated by the
corresponding form of energy.  We define
\begin{mathletters}
\begin{equation}
E=f^n({1 \over 2} {\dot \Phi}^2+V(\Phi)) ,
\label{energy}
\end{equation}
\begin{equation}
T={\rm [period\, of\, oscillation]} = \oint {{d\Phi} \over
{\sqrt{2(f^{-n}E-V(\Phi))}}} .
\label{period}
\end{equation}
\end{mathletters}

Then for periods $<<H^{-1}$, we average the right-hand-side of Eq.
(\ref{edot}) over a period, with the result
\begin{mathletters}
\begin{equation}
\langle 3H(f^n  \dot \Phi )^2 \rangle \simeq (3H/T) \int_0^T f^n
{\dot\Phi}^2 dt =(3H/T) \oint f^n \dot \Phi d\Phi ,
\end{equation}
\begin{equation}
\langle 6M^2H^2  d f^n /dt \rangle \simeq 6M^2H^2 T^{-1} \int_0^T
(df^n/dt) dt \simeq 0 .
\end{equation}
\end{mathletters}
Defining $\gamma = (1/ET) \oint f^n \dot\Phi d\Phi$, we find
\begin{equation}
E^{-1} dE/dt =-3 \gamma H=-3\gamma \dot a /a \,.
\end{equation}

For $\gamma \simeq {\rm constant}$, this equation can be integrated to
give $E \propto a^{-3\gamma}$.  Defining the integrals
\begin{equation}
I_{\pm}=\oint (1-f^n V(\Phi)/E)^{\pm 1/2} f^{n/2} d\Phi ,
\end{equation}
then
\begin{equation}
\gamma= 2I_+/I_- .
\end{equation}
It is useful to write $I_{\pm}$ in terms of $\phi$:
\begin{equation}
I_{\pm} = \oint (1-f^{n-2} \tilde V (\phi) /E)^{\pm 1/2}
(f+\beta)^{1/2} f^{n/2 -1} d\phi .
\end{equation}
Note that when $\Phi$ reaches its extremal values within each
oscillation, $\dot\Phi =0$, so $E=V(\Phi_m)=f(\phi_m)^{-2} \tilde V
(\phi_m)$, where the subscript $m$ denotes the extremal value.  In the
quadratic coupling case, there is a symmetry $\phi \to -\phi$, so the
maximum and minimum values of $\phi$ are identical apart from a sign;
in the linear
coupling case, this is no longer necessarily true.

However, for $\phi$ oscillations generated by inflation, we will find
that $\eta<<1$ already at the onset of oscillations, so that in evaluating
$\gamma$, we can set $f \approx 1$.  We can now treat
$\phi_{max}=-\phi_{min}\equiv \phi_m$ in all cases.  We define
$\zeta_m \equiv \zeta(\phi_m)$, and $x \equiv \phi/\phi_m$.  We also
have, in the
linear case, $\beta=3\alpha^2$ and in the quadratic case,
$\beta_m \equiv \beta(\phi_m)$.  Then, independently of $n$,
\begin{mathletters}
\begin{equation}
{I_{\pm}}^{\rm linear}=4(1+3\alpha^2)^{1/2} \int_0^1 (1-x^2)^{\pm 1/2} (1+
{\zeta_m \over {1+\zeta_m}} x^2)^{\pm 1/2} dx \,,
\end{equation}
\begin{equation}
{I_{\pm}}^{\rm quadratic}=4 \int_0^1 (1-x^2)^{\pm 1/2} (1+{\zeta_m \over
{1+\zeta_m}}  x^2)^{\pm 1/2} (1+\beta_m x^2)^{1/2} dx \,.
\end{equation}
\end{mathletters}
Table \ref{gammatable}
shows the resulting values of $\gamma$ for the appropriate
ranges of $\zeta_m$ and $\beta_m$.   For the linear case and the
quadratic case with small coupling parameter $\beta_m$, $E$ has the expected
variation: as matter ($\propto a^{-3}$) in the harmonic regime, and
as radiation ($\propto a^{-4}$) in the anharmonic regime.  However,
for the quadratic case with large coupling parameter, $E$ for
anharmonic oscillations varies like matter ($\propto a^{-3}$), while for
harmonic oscillations, it decreases more slowly than matter ($\propto
a^{-2}$).  The evolution of the amplitude $\phi_m$ can be obtained from
the fact that $E \approx {1 \over 2} m^2 {\phi_m}^2
(1+\zeta_m)(1+\eta_m)^{n-2}$.  Table \ref{gammatable}
shows the relevant values of
$\delta$, defined by $\phi_m \propto a^{-3\delta}$.

\section{Linearly coupled oscillating G}

\subsection{Oscillations generated by inflation}

Inflation can naturally lead to a finite displacement of $\Phi$ from
zero, leading to finite oscillations.
We assume that a false vacuum epoch occurs during
which $-p=\rho =\rho_V \simeq {\rm constant}$.  Then
$\ddot \Phi+3H \dot\Phi =-dV_I(\Phi)/d\Phi$, where the effective
inflationary scalar potential $V_I$ is given by (Eq. (\ref{Veff}))
\begin{equation}
V_I(\Phi)={{{1 \over 2} m^2 \phi^2 + {1 \over 4} \lambda
\phi^4+\rho_V} \over {(1+\alpha\phi/M)^2}} .
\label{linpotential}
\end{equation}
During the inflationary epoch, the effective damping caused by the $3H
\dot\Phi$ term leaves the scalar field at the minimum of $V_I$, which
occurs at $dV_I/d\Phi=0=dV_I/d\phi$.

Figure \ref{potentialfig}
shows schematically
how the effective potential during inflation, $V_I(\Phi)$,
differs from the effective potential today.
The effect of the nonminimal couplings is to change the shape of
the potential,  shift the minimum away from $\Phi=0$, and raise
the value of the energy density at the minimum.  The
change in shape depends on the  nature of the nonminimal
coupling; {\it e.g.}, compare the upper and lower
panels.
For the linear case, we can
use Eqs. (\ref{y}) and (\ref{Q}),
to obtain the equation for the minimum, expressed in terms of
the amplitude of the oscillations in $f$, $\eta=\alpha\phi/M$,
\begin{equation}
\eta_I (\alpha^2/\lambda y^2 + \eta_I^2 + {1 \over 2} \eta_I^3 ) =
\alpha^4 Q .
\label{minimumlinear}
\end{equation}

The first term in Eq. (\ref{minimumlinear}) comes from $m^2\phi$ in
$dV_I/d\phi$, the second comes from $\lambda \phi^3$, while the third
comes from a cross term between $\lambda \phi^3$ and $\alpha \phi/M$
in the denominator of Eq. (\ref{linpotential}).
Note that, in this case, the anharmonicity parameter $\zeta$
(Eq. (\ref{zeta})), which is essentially the ratio
between the second and first terms on the left-hand-side of
Eq. (\ref{minimumlinear}), can be written
$\zeta={1 \over 2} \lambda y^2 \eta^2 /\alpha^2$.

One class
of solutions has the first term in parentheses in Eq. (\ref{minimumlinear})
dominant;
this corresponds to $\zeta_I<<1$, {\it i.e.} to scalar fields
in the harmonic regime during inflation.  However, because $\rho_\Phi$
falls off as $a^{-3}$ immediately, compared to the $a^{-4}$ fall-off
for radiation, the result can be shown to be
unacceptable over-dominance of the scalar field energy density
today, so we will not
consider this case further (this case corresponds to the lower portion
of Fig. \ref{linbounds}).
The alternative, anharmonic
limit, $\zeta_I>>1$ (second
two terms dominant) yields
\begin{equation}
\eta_I^3 (1+ {1 \over 2} \eta_I) = \alpha^4 Q ,
\label{etaequation}
\end{equation}
with solutions
\begin{equation}
\eta_I \approx
\cases{
\alpha^{4/3} Q^{1/3} << 1 , &{\rm for}\quad $\alpha<<1.5 \times
10^4 \lambda^{1/4}t_V^{-1}$ \,,\cr
2^{1/4} \alpha Q^{1/4} >> 1 , &{\rm for}\quad$\alpha>>1.5 \times 10^4
\lambda^{1/4}t_V^{-1}$ \,.\cr}
\end{equation}
The small $\alpha$ of the first case leads
to a flatter potential (see Fig.
\ref{potentialfig}), with a minimum at a relatively large value of
$\phi$, but a small $\eta$.  The second case leads to a deeper
potential, with a smaller, universal value at the minimum of $\phi/M \approx
(24/\lambda)^{1/4} (\rho_V/6M^4)^{1/4}$, but a large value of $\eta$.
For these solutions, generally $\eta_I /(1+\beta_I )=\eta_I/(1+3\alpha^2)
<<1$ for $\lambda$
and $t_V$ of order unity (see Eq.
(\ref{ratiolinear}) below).

\subsection{Post-inflation evolution of scalar field}

At the end of inflation when reheating occurs, $\dot\Phi \approx 0$,
$\rho-3p=0$, and $\rho_\Phi \approx V(\Phi_I) = ({1 \over 2} m^2
\phi_I^2 + {1 \over 4} \lambda \phi_I^4 )/(1+\alpha\phi_I/M)^2$.  For
simplicity, we assume that the vacuum energy density is totally
converted to radiation energy, so that $\rho_R=\rho_V$.  Since we are
considering the limit
$\zeta_I>>1$ we can approximate $V_I(\Phi) \approx f^{-2}({1 \over 4}
\lambda \phi^4)$.
Then using Eq. (\ref{etaequation}), we find that, right after
inflation ends,
\begin{mathletters}
\begin{equation}
\chi_I \equiv (\rho_\Phi /f^{-2} \rho_R )_I \approx \eta_I^4/2\alpha^4 Q
\approx \eta_I /(2+\eta_I)
,
\label{chisubI}
\end{equation}
\begin{equation}
H^2 \approx (\rho_V/6M^2) (1+\eta_I)^{-1} (1+{1 \over 2} \eta_I )^{-1} .
\end{equation}
\label{hsquared}
\end{mathletters}

However, when the false vacuum energy density is converted to
radiation and $\rho-3p \to 0$, the minimum of the potential $V$ goes
to $\Phi=0$.  Whether $\Phi$ begins to oscillate or rolls slowly down
the potential $V$ depends on whether the period of oscillation is short
or long compared to the expansion time scale, where $({\rm period})^2
\approx \Phi (dV/d\Phi)^{-1} \approx \phi (d\Phi/d\phi)^2
(dV(\Phi)/d\phi)^{-1}$.  In this case, using $d\Phi/d\phi$ (Eq.
(\ref{dphidphi})), we obtain $({\rm period})^2 \approx
m^{-2} (1+\eta_I)(1+\beta+\eta_I)(1+2\zeta_I+\eta_I\zeta_I)^{-1}$.  Since
we have argued that $\eta_I/(1+\beta)<<1$, where $\beta=3\alpha^2$,
and $\zeta_I>>1$, we
then find the ratio
\begin{equation}
{\cal R} \equiv {({\rm period})^2 \over H^{-2}} \approx
\left ( {{1+3\alpha^2} \over {6\alpha^2}} \right )
\left ( {{\eta_I} \over {2+\eta_I}} \right ) .
\label{ratiolinear}
\end{equation}
For the solution with $\eta_I<<1$, ${\cal R} << 1$, so oscillations
begin immediately, but with sufficiently small amplitude that we can
approximate $f \approx 1$ right away.  Then, initially, $\rho_\Phi
\approx E \propto a^{-4}$, since the oscillations are anharmonic, and
$\phi_m \propto a^{-1}$.  Since $\zeta_m \propto \phi_m^2 \propto a^{-2}$,
the transition to harmonic oscillations (H) at $\zeta_m \approx 1$
occurs at $a_H \approx \zeta_I^{1/2} a_I$.  In the harmonic regime,
$\rho_\Phi \propto a^{-3}$ and $\phi_m \propto a^{-3/2}$.  It then
follows
that the
present ratio of scalar to radiation energy densities is given by
\begin{equation}
\chi_0 = (\rho_\Phi/f^{-2} \rho_R)_0 \approx \chi_I \zeta_I^{-1/2}
(a_0/a_I) \approx (2\lambda)^{-1/2} (\alpha/y)(a_0/a_I) ,
\label{chitoday}
\end{equation}
and the present amplitude of G-oscillations is $\eta_0 \approx \eta_I
\zeta_I^{1/2} (a_I/a_0)^{3/2}$.  Substituting Eq. (\ref{scalefac})
for $a_0/a_I$
into Eq. (\ref{chitoday}), we obtain for the present scalar density parameter
\begin{equation}
\Omega_\Phi^0 h^2= \Omega_R^0 h^2 \chi_0 \approx 10^{22} (2/\lambda)^{1/2}
(\alpha/y)
(T_V/10^{14} {\rm GeV}) ,
\end{equation}
where we have used Eq. (\ref{omegarad}) for the radiation density
parameter and have assumed $T_0=2.73^{\rm o}$ K.
If G-oscillations are to account for
the energy density needed to close the universe without other forms of
dark matter, then we require $\Omega_{\Phi}^0 \approx 1$.
For dimensionless coupling
$\lambda$ of order unity and $T_V \approx 10^{14} {\rm GeV}$,
the values of the
scalar mass $m=M/y$ and the coupling constant $\alpha$ that will meet
this constraint are shown in
Fig. \ref{linbounds}.
If the universe
is flat, then $h<.65$ is needed to satisfy age constraints from globular
cluster ages.  In Fig.~2, the solid dark line corresponds to
$\Omega_{\Phi}^0 h^2=0.25$,
{\it e.g.}, $\Omega_{\Phi}^0=1$ and $h=0.5$.
The curve shows that $\Omega_{\Phi}^0$ near
closure density is possible for a wide range of parameters, including
the range where dimensionless parameters, such as $\lambda$ and $\alpha$
of are of order unity.
Notice in Fig. \ref{linbounds}
that the present amplitudes of
$G$-oscillation are extremely small, despite contributing a potentially
critical
energy density.   Also plotted is the curve for $\Omega_\Phi^0 h^2=1$
(dashed curve, parallel to dark curve), which is, roughly, the upper
bound on $\Omega_{\Phi}^0 h^2$ consistent with observations.

For the case $\eta_I>>1$, corresponding to the right-hand strip of
Fig. \ref{linbounds}, the scalar energy density is comparable to the
radiation energy density just following inflation (Eq. (\ref{chisubI})),
and the ratio $\cal
R$ (Eq. (\ref{hsquared})) is of
order unity.  As a result, oscillations do not start
immediately, instead, the scalar field rolls slowly down the
potential, while the radiation energy density falls off rapidly.
Eventually, $\eta$ (and thus $\cal R$) becomes small enough that
oscillations commence, first anharmonically, and then harmonically as
before.  But because of the intervening decrease in the radiation
energy density, the resulting scalar energy density overdominates the
universe today.  This effect causes the curves of constant
$\Omega_\Phi^0 h^2$ in Fig. \ref{linbounds}
to turn up as they approach the line where
$\eta_I \approx 1$ ($\alpha \approx 10^4 \lambda^{1/4}t_V^{-1}$).

\section{Quadratically coupled oscillating G}

\subsection{Oscillations generated by inflation}

In the quadratically coupled case, the effective inflationary scalar
potential (Eq. (\ref{Veff})) is given by
\begin{equation}
V_I(\Phi)={{{1 \over 2} m^2 \phi^2 + {1 \over 4} \lambda
\phi^4+\rho_V} \over {(1+\xi\phi^2/M^2)^2}} .
\end{equation}
Note that as $\phi \to \infty$, the potential tends to a constant,
$\lambda M^4/4\xi^2$.   Fig. \ref{potentialfig}
plots $V_I(\Phi)$ schematically.  The
minimum of the potential occurs at
\begin{equation}
\eta_I =2\xi^2 Q r
\cases{
<< 1 , &$\xi<<10^8 \lambda^{1/2}t_V^{-2}r^{-1/2}$\,,\cr
>> 1 , &$\xi>>10^8 \lambda^{1/2}t_V^{-2}r^{-1/2}$\,,\cr}
\label{minimumquad}
\end{equation}
where $Q$ is given by Eq. (\ref{Q}), and
\begin{equation}
r \equiv {{1-[24\xi y^2 \rho_V/6M^4]^{-1}} \over {1-\xi y^{-2}
\lambda^{-1}}} >0 .
\label{r}
\end{equation}
If $r<0$, the minimum occurs at $\eta_I=0$, and no
oscillations are generated by inflation.  If $r>0$, but both numerator
and denominator in Eq. (\ref{r}) are negative, then $\eta_I=2\xi^2Qr$
is a maximum, while $\eta_I =0$ is a minimum, again resulting in no
oscillations.  These cases correspond to the region labelled ``no
oscillations'' in Fig. \ref{quadbounds}.
For the quadratically coupled model, the anharmonicity parameter is
given by $\zeta={1 \over 2}\lambda y^2\eta/\xi$.
Note that for oscillations to occur, the numerator of Eq. (\ref{r})
must be positive, thus
$24\xi y^2(H_I/M)^2 = \zeta_I/r >1$, corresponding to the anharmonic
regime.  The larger $\zeta_I$ is, the closer $r$ is to unity.

\subsection{Post-inflation evolution of scalar field}

In this case, we find at the end of inflation,
\begin{mathletters}
\begin{equation}
\chi_I  \approx \eta_I^2/2\xi^2 Q \approx \eta_I  ,
\end{equation}
\begin{equation}
H^2 \approx (\rho_V/6M^2) (1+\eta_Ir)(1+ \eta_I )^{-2} .
\end{equation}
\end{mathletters}
The effective oscillation period is $({\rm period})^2 \approx m^{-2}
(1+\eta_I)(1+\beta_I+\eta_I)(1+2\zeta_I-\eta_I)^{-1}$, Using Eq. (\ref
{minimumquad}) and the fact that $\zeta_I>>1$, we find that the ratio
$\cal R$ is
\begin{equation}
{\cal R}\approx {{1+\beta_I+\eta_I} \over {24\xi r}}\approx
{{1+(12\xi+1)\eta_I} \over {24\xi}} .
\end{equation}
There are three possible post-inflation evolutions.

{\it (a) Immediate Oscillations.}
Oscillations begin immediately when ${\cal R}<<1$, which corresponds to
$1/24 <\xi<2.9 \times 10^7 \lambda^{1/2}$.   In this range,
$\eta_I<<1$,
so we approximate $f \approx 1$, and $\rho_\Phi << \rho_R$.  However,
although
$\zeta_I>>1$, $\beta_I=12\xi \eta_I$ ranges from $10^{-18}$ to $10^{11}$, with
$\beta_I \approx 1$ for $\xi \approx 3 \times 10^5 \lambda^{1/3}
t_V^{-4/3} r^{-1/3}$, so we must consider two separate cases (for the
range of parameters considered, $\beta_I/\zeta_I \approx 24\xi^2/y^2
<<1$ ).  Recall that $\beta={3 \over 2} (1+\eta)/\omega$, so that
large $\beta$ corresponds roughly to small $\omega$ and to strong
Brans-Dicke-like effects (we call this ``strong non minimal coupling''
or ``strong coupling'') and small
$\beta$ corresponds to large $\omega$ and to weak Brans-Dicke effects
(we call this ``weak coupling'').

{\it Case 1:} $\beta_I>>1$: Because of the strong
non-minimal coupling,
the energy density falls
off like matter rather than like radiation despite the anharmonic
potential (Table \ref{gammatable}), so that
initially $E \approx \rho_\Phi
\propto a^{-3}$, while $\beta_m ,\> \zeta_m, \> \eta_m \propto
a^{-3/2}$;
The transition to weak coupling (WC)
($\beta_m \approx 1$) occurs at $a_{WC} \approx \beta_I^{2/3} a_I$,
whereupon standard anharmonic oscillations lead to
$E \propto a^{-4}$ and $\beta_m ,\> \zeta_m, \> \eta_m
\propto a^{-2}$.  Harmonic oscillations commence at $a_H \approx
\zeta_{WC}^{1/2} a_{WC} \approx \zeta_I^{1/2} \beta_I^{1/6} a_I$.
During harmonic oscillations, $E \propto a^{-3}$ and $\beta_m ,\>
\zeta_m, \> \eta_m \propto a^{-3}$.   The result at the present epoch
is
\begin{mathletters}
\begin{equation}
\chi_0 \approx \chi_I \zeta_I^{-1/2} \beta_I^{1/2} (a_0/a_I) \approx
\eta_I \zeta_I^{-1/2} \beta_I^{1/2} (a_0/a_I) ,
\end{equation}
\begin{equation}
\eta_0 \approx \eta_I \zeta_I^{1/2} \beta_I^{1/2} (a_I/a_0)^3 ,
\end{equation}
\end{mathletters}
and
\begin{equation}
\Omega_\Phi^0 h^2 \approx 4\times 10^6 \lambda^{-3/2} (\xi^3/y)
(T_V/10^{14} {\rm GeV})^5 .
\label{quadbound1}
\end{equation}
Note that, for larger values of $\xi$ within the range quoted above,
namely $7 \times 10^5 \lambda^{5/13}<\xi<2.9 \times 10^7 \lambda^{1/2}$
the scalar energy density can actually surpass the radiation energy
density during the strong-coupling regime.  For the smaller values of
$\xi$, the scalar density approaches but does not exceed the radiation
during the strong-coupling regime; it is only when harmonic
oscillations begin that the scalar energy surpasses the radiation.
These two sub-cases are shown as curves (a) and (b) in Fig.
\ref{falloff}.

{\it Case 2:} $\beta_I<<1$:  In this case, weak coupling holds right
after inflation.
Standard anharmonic oscillations begin
immediately, $E \propto a^{-4}$ and $\beta_m ,\> \zeta_m, \> \eta_m
\propto a^{-2}$ , with harmonic oscillations taking over when $a_H
\approx \zeta_I^{1/2} a_I$ (curve (c) in Fig. \ref{falloff}).
The evolution then proceeds as in the previous
case, with the result
\begin{mathletters}
\begin{equation}
\chi_0 \approx \eta_I \zeta_I^{-1/2} (a_0/a_I) ,
\end{equation}
\begin{equation}
\eta_0 \approx \eta_I \zeta_I^{1/2} (a_I/a_0)^3 ,
\end{equation}
\end{mathletters}
and
\begin{equation}
\Omega_\Phi^0 h^2 \approx 1.6\times 10^{14} \lambda^{-1}
(\xi^{3/2}/y)(T_V/10^{14} {\rm GeV})^3 .
\label{quadbound2}
\end{equation}

{\it (b) Deferred oscillations}.  For $\xi << 1/24$,
$\eta_I<<10^{-17}$, $\chi_I<<1$, but ${\cal R}>>1$ and $\zeta_I>>1$.
This corresponds to small amplitudes for $\phi$, though still in the
anharmonic regime, but since $m/H<<1$ at this epoch, the scalar field
first rolls slowly down the potential.
Thus we can ignore
$\ddot \Phi$ in Eq. (\ref{cosmo-b}), and solve the approximate ``slow-roll''
evolution equation
\begin{equation}
\dot \phi = (d\phi/d\Phi) \dot \Phi \approx -(3H)^{-1} (d\phi/d\Phi)^2
dV(\Phi)/d\phi \approx - (m^2 \phi^2/3H)(1+2\zeta).
\label{slowroll1}
\end{equation}
Radiation strongly dominates the energy density following inflation,
so $H \sim 1/2t$.  Noting that $\zeta \propto \phi^2$, we rewrite Eq.
(\ref{slowroll1}) in the form $\dot \zeta \approx -{4 \over 3} m^2 t
\zeta (1+2\zeta)$.  For $mt \sim m/H<<1$ we find the approximate solution
$\zeta \approx \zeta_I /(1+{4 \over 3}
\zeta_I m^2 t^2)$.  Noting that $m^2 \zeta_I \approx 12\xi
(\rho_V/6M^2)$ and defining $\tau \equiv t(\rho_V/6M^2)^{1/2}$, we
obtain
\begin{equation}
{\cal R} \approx (2t)^{-1} (2m^2 \zeta)^{-1} \approx [1+16\xi
\tau^2]/96\xi\tau^2 .
\end{equation}
Thus oscillations begin when $\tau \approx (80\xi)^{-1/2}$.  At this
point, $\rho_R \approx 2 \rho_I \tau^{-2} \approx 20 \xi \rho_I$,
$\rho_\Phi \approx \lambda\phi^4 \approx  (25/36) \rho_\Phi^I $, and
$a \approx a_I/(20\xi)^{1/4}$.  Following the oscillations from this
point onward, we find
\begin{equation}
\Omega_\Phi^0 h^2 \approx 10^{13} \lambda^{-1} (\xi^{3/4}/y)
(T_V/10^{14} {\rm GeV})^3 .
\label{quadbound3}
\end{equation}
This corresponds to curve (d) in Fig. \ref{falloff}.

{\it (c) Lingering inflation}.  For $\xi >> 3\times 10^8
\lambda^{1/2}$, $\eta_I>>1$, we have $\rho_\Phi \approx
(\lambda/4\xi^2)M^4$, $\chi_I \approx \eta_I >> 1$, and ${\cal R}
\approx \eta_I /2 >>1$.  In this case,  the scalar-field energy
density dominates, and the scalar field rolls slowly down the
flat wing of the
post-inflation potential (Figure \ref{potentialfig}).  In
this case, the slow-roll evolution equation gives
\begin{equation}
\dot \phi \approx -(3H)^{-1} (d\phi/d\Phi)^2
dV(\Phi)/d\phi \approx -{1 \over 3} (\lambda/6)^{1/2} \xi^{-2} M^3
\phi^{-1} ,
\end{equation}
where we have used $H \approx (\lambda/24)^{1/2} \xi^{-1} M$, and
$\beta, \> \eta , \> \zeta >>1$.  The solution is $\eta_I -\eta
\approx {1 \over 3} (\lambda/6)^{1/2} \xi^{-1} M (t-t_I)$.
Oscillations in $\Phi$ commence when $\eta \approx 1 <<\eta_I$, {\it
i.e.} when $H (t-t_I) \approx {3 \over 2} \eta_I$.  During this
period, $H \approx {\rm constant}$, $a \propto e^{Ht}$, and the
radiation density deflates by $e^{-4H(t-t_I)} \approx e^{-6\eta_I}$.  This is
a kind of ``lingering inflation'', driven by the approximately constant
scalar energy density on the flat portion of the potential at large
$\phi$ (Figure \ref{potentialfig}).
Eventually oscillations begin and the evolution
proceeds as in the previous case, except that the lingering inflation
leads to an effective initial $\chi_I$ larger than before
by the factor $e^{6\eta_I} >>1$.
The result is a much larger value for the scalar-field energy
density:
\begin{equation}
\Omega_\Phi^0 h^2 \approx 4\times 10^6 \lambda^{-3/2} (\xi^3/y)
e^{(4\times 10^{-16} \lambda^{-1}\xi^2 t_V^4)} (T_V/10^{14} {\rm
GeV})^5 .
\label{quadbound4}
\end{equation}
Because $\xi>3\times 10^8 \lambda^{1/2}$
in this case, we see that the scalar field
overdominates the present energy density by many orders of magnitude.
This range of values of $\xi$ is therefore unviable.

As discussed for the linear case, $\Omega_\Phi^0 h^2 \approx 0.25$
corresponds to the case where the oscillation energy in $\Phi$ is
near the closure density, assuming no additional dark matter.
Imposing $\Omega_\Phi^0 h^2 \approx 0.25$ and assuming $\lambda$ and
$T_V/10^{14} {\rm GeV}$ of order
unity, we obtain from
Eqs. (\ref{quadbound1}),
(\ref{quadbound2}),
(\ref{quadbound3}), and
(\ref{quadbound4})
the constraint on the scalar mass $m=M/y$ and the
coupling constant $\xi$ shown in Fig. \ref{quadbounds}.
As with the linear case, we find that
$\Omega_{\Phi}^0$ near
closure density is possible for a wide range of parameters, including
the range where dimensionless parameters, such as $\lambda$ and $\xi$
of are of order unity.
Also shown is the corresponding curve for
$\Omega_\Phi^0 h^2 = 1$, roughly the largest value compatible with
observations..
Not surprisingly, given
the quadratic nature of the non-minimal coupling, the amplitude of
G-oscillations at the current epoch is extraordinarily small, despite
their ability to generate a closure energy density.

\section{Constraints from local experiments}

We turn now to a discussion of bounds on these models imposed by local
laboratory and solar-system experiments.  Because of the presence of
the potential $V(\Phi)$, there is a length scale $\ell$ over which
modifications to general relativity may be important.  For scales much
greater than $\ell$, the theory is equivalent to general relativity.
Numerous tests of general relativity and of the inverse-square law of
gravity using a variety of
techniques have been performed which cover a range of lengths from
solar-system scales down to centimeters.  In this section we present a
general method for using the results of such experiments to bound
the parameters of oscillating $G$ theories.  We will find that the
most viable models from a cosmological viewpoint are comfortably
consistent with all local experimental limits.

Comparison with experiment is most easily done in the physical
representation.  We expand the physical metric and the scalar field
about their asymptotic values, $\tilde g_{\mu\nu}
=\eta_{\mu\nu} + h_{\mu\nu}$, and $\phi = \phi_0 (t)+\varphi$, and
write $f(\phi)=f_0+f_0^\prime \varphi + {1 \over 2}
f_0^{\prime\prime} \varphi^2 $, and substitute into Eqs. (\ref{allfieldeq}),
keeping
terms to first order in $h_{\mu\nu}$ and $\varphi$.  In eliminating
the zeroth order terms in the field equation for $\phi$, corresponding
to the cosmological solution for $\phi_0$, we note that the coordinate
$t$ used here is related to the time $t_{\rm RW}$ of the physical
Robertson-Walker metric by $dt_{\rm RW}=(1-{1 \over 2}h_{00})dt$.  The
result is
\begin{equation}
-\ddot \varphi+(\nabla^2-\tilde m^2)\varphi =-{1 \over 2} {{f_0^\prime}
\over {f_0+3M^2f_0^{\prime 2}}} (\rho-3p)
+{1 \over 2} {{f_0^\prime (1+6M^2f_0^{\prime\prime})} \over
{f_0+3M^2f_0^{\prime 2}}} \dot\phi_0 \dot\varphi +{1 \over 2} \dot
h_{ii} \dot\phi_0 ,
\end{equation}
where
\begin{eqnarray}
\tilde m^2 &= & {{m^2} \over {f_0+3M^2f_0^{\prime 2}}}  \biggl\{
 f_0-f_0^\prime
\phi_0 - f_0^{\prime\prime} \phi_0^2 - {{f_0^\prime \phi_0 (f_0-f_0^\prime
\phi_0)(1+6M^2f_0^{\prime\prime})} \over {f_0+3M^2f_0^{\prime 2}}}
 \nonumber\\
&&   -{1 \over 2} {{\dot\phi^2} \over {m^2}} {{1+6M^2f_0^{\prime\prime}}
\over {f_0+3M^2f_0^{\prime 2}}} (f_0^{\prime 2} -
f_0f_0^{\prime\prime} +3M^2f_0^{\prime 2}f_0^{\prime\prime})
\biggr\} \, . \label{mtilde}
\end{eqnarray}

We are interested in approximately static solutions corresponding to a
gravitating mass such as the Sun or Earth, approximately at rest ($v
\approx 0 \,, p<<\rho$).
Thus we can drop the terms involving time derivatives.  Using Eq.
(\ref{omega}), and averaging the oscillations of $\phi_0$ over several
periods, we find
\begin{equation}
(\nabla^2-\tilde m^2)\varphi = -\left ({{f_0^\prime} \over f_0} \right
) {\omega \over
{(3+2\omega)}} \rho ,
\end{equation}
whose solution is
\begin{equation}
\varphi = 4 M^2 f_0^\prime {\mu \over r}
e^{-\tilde m r} ,
\end{equation}
where $\mu = f_0^{-1} GM_{\rm source}$ is the effective gravitational
mass of the gravitating body.

Using the approximations $f_{;00} \approx 0$, $\Box f \approx
f_0^\prime \nabla^2 \varphi$, $T_{00} \approx \rho$, $T_{ij} \approx 0$,
and working in a gauge in which ${h_i^\mu}_{,\mu} - {1 \over 2}
{h_\mu^\mu}_{,i} = (f_0^\prime /f_0)\varphi_{,i}$, we obtain the field
equations
\begin{mathletters}
\label{del2h}
\begin{equation}
\nabla^2 (h_{00} - (f_0^\prime /f_0) \varphi) =-\rho/2M^2f_0 -
(\rho_\phi +3p_\phi)/2M^2f_0 ,
\end{equation}
\begin{equation}
\nabla^2 (h_{ij} + (f_0^\prime /f_0) \delta_{ij} \varphi) =-\delta_{ij}
\rho/2M^2f_0 - \delta_{ij} (\rho_\phi - p_\phi)/2M^2f_0 ,
\end{equation}
\end{mathletters}
where $\rho_\phi = {1 \over 2} \dot\phi_0^2 +\tilde V (\phi_0)$ and
$p_\phi = {1 \over 2} \dot\phi_0^2 -\tilde V (\phi_0)$.  Dropping the
$\rho_\phi$ and $p_\phi$ terms for the moment, we obtain the
solutions
\begin{mathletters}
\label{hmunu}
\begin{equation}
h_{00} = {{2\mu} \over {r}} \left ( 1+ {1 \over
{3+2\omega}}e^{-\tilde m r} \right ) ,
\end{equation}
\begin{equation}
h_{ij} = {{2\mu} \over {r}} \delta_{ij} \left ( 1- {1 \over
{3+2\omega}}e^{-\tilde m r} \right ) .
\end{equation}
\end{mathletters}
The effective mass $\tilde m$ leads to a range $\tilde m^{-1}$ for
modifications of Einstein gravity.  The ``Newtonian'' potential
$h_{00}$ can be written in the more familiar form $h_{00} = 2G_{eff}\mu/r$,
where $G_{eff}$ has the effective form
\begin{equation}
G_{eff} = \left ( 1+ {1 \over {3+2\omega}}e^{- \tilde m r} \right ) .
\label{Geff}
\end{equation}
Because of the Yukawa-type correction to the potential, for $r<<\tilde
m^{-1}$, $G_{eff} \approx (4+2\omega)/(3+2\omega)$, while for $r>>\tilde
m^{-1}$, $G_{eff} \approx 1$.  Thus, for $r>>\tilde
m^{-1}$, the theory is equivalent to general relativity.  Experimental
tests of classical general relativity can in principle
be used to place bounds on
$\tilde m$ and on the coupling parameters $\alpha$ and $\xi$ of the
two oscillating $G$ models.

\subsection{Light deflection and time delay}

To lowest order in $\mu/r$, the physical metric obtained from Eqs.
(\ref{hmunu})
can be written in the form
\begin{mathletters}
\label{gmunu}
\begin{equation}
g_{00} \approx -(1-{{2\mu} \over r})(1-{{2\mu} \over {3+2\omega}}
{{e^{-\tilde m r}} \over r}) ,
\end{equation}
\begin{equation}
g_{ij} \approx  \delta_{ij} (1+{{2\mu} \over r})(1-{{2\mu} \over {3+2\omega}}
{{e^{-\tilde m r}} \over r}) .
\end{equation}
\end{mathletters}
Notice that, in the line element, $ds^2=g_{\mu\nu}dx^\mu dx^\nu$, the
exponential terms can be extracted as an overall conformal factor.
Hence, there is no direct effect on null geodesics, and the deflection
and time delay of a signal is proportional to $\mu$, with no explicit
dependence on $\tilde m$.  However, $\mu$ represents the mass as
measured at spatial infinity using Kepler's law.  In solar system
applications, such as the deflection of light, the solar mass actually
used is measured by using Kepler's law applied to the Earth's orbit.
If the range $\tilde m^{-1}$ is small compared to the Earth's orbit,
{\it i.e.} $\tilde m^{-1} <<10^8 {\rm km}$, corresponding to $\tilde m
>> 10^{-27}$ GeV, or an oscillation frequency $\nu >> 10^{-2}$ Hz, the inferred
solar mass
$M_\odot$ is the same as $ \mu$ in Eq. (\ref{gmunu}), and consequently, the
predicted time delay and light deflection  in terms of the measured
solar mass are indistinguishable from Einstein relativity.  This
corresponds to $y << 10^{45}$, which easily includes the masses of
interest for cosmology.
Conversely, if $\tilde m^{-1}>>10^8$ km, the inferred mass is $M_\odot
=\mu (4+2\omega)/(3+2\omega)$, and the prediction for time-delay and
light deflection approaches the Brans-Dicke form, proportional to
$(3+2\omega)/(4+2\omega)$.

\subsection{Orbits}

Using Earth-LAGEOS-Lunar data \cite{rapp} which compares $GM_\oplus$
inferred from Earth-surface gravity measurements to the values
inferred from the study of the orbits of the LAGEOS satellites and the
Moon, tight bounds can be obtained on the variation in $G$ in the
$10^4$ -- $10^8$ m range.  From a rough fit to the
appropriate portion of Fig. 1
in \cite{fifthforce}, we find that the coefficient in the
exponential correction in Eq. (\ref{Geff}) is constrained to be
\begin{equation}
(3+2\omega)^{-1}<0.03 (\nu/{\rm MHz})^{1.1},
\end{equation}
for frequencies in the 10 kHz to 1 Hz range, corresponding to values of
$y$ between $10^{38}$ and $10^{42}$.   For the higher frequencies of
interest to our cosmological models, the bound is easily satisfied.

\subsection{Tower and laboratory measurements}

Tower measurements of the variation in $G$ place the limit
$(3+2\omega)^{-1}<2\times 10^{-4}$ at a range of 100 meters {\it i.e.} $\nu =
3$ MHz, or $y= 10^{36}$.  Stronger limits are obtained from laboratory
tests.  From a null test of Gauss's law at 1.5 meters using
superconducting gravity gradiometers \cite{paik} ($\nu=200$ MHz,
$y=10^{34}$), we find the bound
\begin{equation}
(3+2\omega)^{-1}<3 \times 10^{-5} \ell^2
e^{3/\ell},
\end{equation}
where $\ell$ is the range in meters.  From a test
of the inverse square law at 2 centimeters using a torsion balance
\cite{hoskins} ($\nu=15$ GHz, $y=10^{32}$), we obtain
\begin{equation}
(3+2\omega)^{-1}< 3.4 \times 10^{-2}
\ell^2 e^{0.04/\ell}.
\label{torsion}
\end{equation}
The bounds relax exponentially with increasing mass and
frequency (decreasing range), and quadratically with decreasing mass
and frequency (increasing range).
Notice that the conversion from range
$\ell$ to the variable $y$ is $\ell/1 {\rm meter} = 1.14 \times
10^{-34} y (m/\tilde m)$.  Since the range of values of $y$ of
interest to cosmology is $y<10^{30}$, only the bound from the
shortest-range experiment, Eq. (\ref{torsion}) will be meaningful.

\subsection{Application to specific models}

{\it Linear Coupling.}  With small amplitude oscillations at the
present epoch ($\eta_0 <<1$), we find from Eqs. (\ref{omega}) and
(\ref{mtilde})
\begin{equation}
\omega=1/2\alpha^2 \,,\,\,\, \tilde m^2 =m^2/(1+3\alpha^2) .
\end{equation}
For $y<10^{30}$, Eq. (\ref{torsion})
yields the bound $\alpha y<4 \times 10^{31}$.
This bound appears in the upper right corner of Fig. \ref{linbounds}.

{\it Quadratic Coupling.} In this case, to first order in $\eta_0$ and
$\beta_0$, we have
\begin{equation}
\omega=3/2\beta_0 \,,\,\,\, \tilde m^2 =m^2 (1-5\eta_0-2\beta_0) .
\end{equation}
But because $\eta_0$ and $\beta_0$ are so small (see Fig. \ref{quadbounds}),
$\omega$
is sufficiently large to be compatible with any experimental bounds;
in this case, despite the massive scalar field, general relativity is
a strong attractor at the present epoch.

\subsection{Effective cosmological constant}

We now consider the terms in Eqs. (\ref{del2h}) that involve the energy density
and pressure of the oscillating scalar field.  For harmonic
oscillations, averaged over a period, $\langle \rho_\phi \rangle \approx {\rm
constant}$, varying only on a
Hubble timescale, while $\langle p_\phi \rangle \approx 0$.  The resulting
spatially uniform cosmological
scalar energy density acts like an effective cosmological constant,
resulting in contributions to the local physical metric of the form
\begin{mathletters}
\begin{equation}
g_{00} \approx -1 - \Omega_\phi^0 H_0^2 r^2 \,,
\end{equation}
\begin{equation}
g_{ij} \approx \delta_{ij} (1 - \Omega_\phi^0 H_0^2 r^2 ) \,.
\end{equation}
\end{mathletters}
These terms will modify the gravitational dynamics of local systems,
ranging from the solar system to clusters of galaxies.  In the case of
the usual cosmological constant $\Lambda$, observational bounds from
such systems are generally not sufficient to constraint the energy
density represented by $\Lambda$ to values below the critical energy
density \cite{kolbturner}.  Thus those observations are consistent with our
desire to have $\Omega_\Phi^0  \approx 1$.

\section{Conclusions}

The scenario we have investigated in this paper links together several
plausible components.  First, we have assumed some modification of
standard, Einstein gravity in the form of a non-minimally coupled
 scalar field.  Modifications are required in almost
any attempt to obtain a unified quantum theory
 of  particle physics and general relativity, including supergravity and
superstring theories. The specific non-minimal coupling considered
in this paper is similar to the coupling of the dilaton.  Second, we have
assumed that the non-minimally coupled field is massive and has typical
self-interactions of quantum scalar fields.  Third, we have assumed
inflation, which is well-motivated for many independent reasons.

We have shown that, taken together, these components lead naturally to a
scenario in which $G$ oscillates about its mean value. Inflation shifts
the scalar field away from its minimum.  At
some time after inflation, the field
begins to oscillate.   If the mass of the
scalar field is typical of elementary particles, $m \ge 1$~GeV or so, the
oscillations are high frequency compared to the expansion rate of the
Universe.   The amplitude is exponentially small, though, since
the Hubble expansion steadily redshifts the kinetic energy density between
inflation and the present epoch.
For a wide range of model
parameters, the combination of the high frequency and small amplitude
that typically arises passes all known tests from cosmology and
general relativity.


Our analysis applies to models in which
 the $\Phi$ field evolves purely according
to the classical equations of motion (\ref{cosmo-b}) and  the oscillation
energy dissipates through Hubble red shift.  Since the
$\Phi$-field  is nonminimally coupled,
it is  possible in many theories
for the $\Phi$-field  energy to dissipate also  through
decay into other particles via  gravitational interactions.
Decay commences  once the age
 of the Universe exceeds the lifetime
of the $\Phi$-field.  The  classical calculations
leading to Figs.~2 and~3
significantly overestimate the
oscillation amplitude and energy today.
The possibilities depend upon the details of the model and lie beyond
the scope of this paper.

The provocative possibility raised by the analysis is that the oscillation
energy in $G$ could account for a large fraction of the energy density
needed to close the Universe.  Here, even the most extreme case,
where oscillation energy is the only nonluminous energy density,
has been shown to be
viable for a wide range of parameters, including the range for which
dimensionless parameters are of order unity.
A future challenge is to determine if oscillation energy by itself
or combined with conventional dark matter can account for
large-scale structure formation.  Perhaps these investigations will
motivated new experimental approaches for measuring high frequency
oscillations of Newton's constant, or other local effects of such
theories.

\acknowledgments

This research was supported in part by the Department of
Energy Grant DOE-EY-76-C-02-3071 at the  University of
Pennsylvania, and by the National Science Foundation, Grant Nos.
89-22140 and 92-22902 at Washington University.

\newpage
\begin{figure}
\caption{
Schematic plot of effective inflationary scalar potential
during and after inflation}
\label{potentialfig}
\end{figure}

\newpage
\begin{figure}
\caption{
Allowed values of $y=M/m$ and $\alpha$ in the linearly
coupled model.  Hatched line corresponds to bounds from laboratory
inverse square-law tests.  The solid dark line corresponds to
$\Omega_\Phi^0 h^2 =0.25$,  the case of  a critical density of
scalar field energy today and $h=0.5$.
Also shown is $\Omega_\Phi^0 h^2>1$ (dashed line and shading),
roughly the upper limit consistent
with present observations.
Dotted lines are lines of constant amplitude of present-day $G$-oscillations
($\eta_0$). Also shown (dashed lines with shading)
are the excluded regions in which
$\eta_I\gg1$ or $\zeta_I \ll 1$.}
\label{linbounds}
\end{figure}

\newpage
\begin{figure}
\caption{
Allowed values of $y=M/m$ and $\xi$ in the quadratically
coupled model.  Here, the oscillation amplitude ($\eta_0$) is so tiny
that
 laboratory experiments provide no useful bounds.
The excluded regions are where the field is not displaced from the
minimum of the effective potential during inflation (bottom shaded
region) or where there is lingering inflation due to the $\Phi$ field
such that oscillation energy ultimately overdominates the universe.
}
\label{quadbounds}
\end{figure}
\newpage
\begin{figure}
\caption{
Schematic evolution of scalar energy density, $\rho f^2$,
 in quadratically
coupled model for various ranges of coupling parameter $\xi$ and mass
$m$, all leading to the same scalar energy density today. Superimposed
are solid curves showing the evolution of the baryon density, $\rho_B$,
and the radiation energy density, $\rho_R$.   The scalar
energy density corresponds to the various dashed curves:  Curves
(a) and (b): Oscillations begin immediately after inflation, but in
the strong nonminimally coupled regime, where
$\rho f^2 \propto a^{-3}$.  Depending on value of $\xi$ scalar energy
density may grow to exceed that of the  radiation density,  {\it e.g.},
curve (a), before entering the weakly-coupled regime where
$\rho f^2$ begins to evolve as $a^{-4}$.
Curve  (c):
Immediate oscillations in weakly-coupled regime fall off as radiation
from the start.   Curve (d): Deferred oscillations cause scalar energy
density to remain roughly constant initially, then to fall off like
radiation.  Eventually harmonic oscillations take over in all cases,
leading to matter-like falloff. In all cases, the late, weakly-coupled
behavior switches to harmonic oscillations with $\rho f^2 \propto a^{-3}$.}
\label{falloff}
\end{figure}

\newpage
\begin{table}
\caption{Values of $\gamma$ and $\delta$ for scalar-field oscillations varying
according to $E \propto a^{-3\gamma}$ and $\phi_m \propto
a^{-3\delta}$}
\label{gammatable}
\begin{tabular}{lccc}
&Linear&\multicolumn{2}{c}{Quadratic}\\
&&$\beta_m<<1$&$\beta_m>>1$\\
\tableline
\multicolumn{4}{c}{Values of $\gamma$}\\
\tableline
Harmonic $\zeta_m<<1$&1&1&2/3\\
Anharmonic $\zeta_m>>1$&4/3&4/3&1\\
\tableline
\multicolumn{4}{c}{Values of $\delta$}\\
\tableline
Harmonic $\zeta_m<<1$&1/2&1/2&1/3\\
Anharmonic $\zeta_m>>1$&1/3&1/3&1/4\\
\end{tabular}
\label{table1}
\end{table}
\end{document}